\documentclass[conference,10pt]{IEEEtran}
\usepackage{cite}

\usepackage{graphicx}
\usepackage{ifpdf}
\ifpdf
  \usepackage{epstopdf}
\fi

\usepackage[cmex10]{amsmath}
\usepackage{amsthm}
\usepackage{amssymb}
\usepackage{cases}
\usepackage{bm}

\usepackage[caption=false,font=footnotesize]{subfig}

\usepackage{fixltx2e}

\usepackage{url}

\begin{document}
\title{Seeing the Unobservable: Channel Learning for Wireless Communication Networks}
\author{\IEEEauthorblockN{Jingchu Liu, Ruichen Deng, Sheng Zhou, and Zhisheng Niu}
        \IEEEauthorblockA{Tsinghua National Laboratory for Information Science and Technology\\
        Department of Electronic Engineering, Tsinghua University\\
        Beijing 100084, China \\
        Email: \{liu-jc12,drc09\}@mails.tsinghua.edu.cn,  \{sheng.zhou,niuzhs\}@tsinghua.edu.cn\\}}

\maketitle
\theoremstyle{plain}
\newtheorem{theo:exist}{Theorem}
\theoremstyle{remark}
\newtheorem{dis:ass1}{Discussion}
\newtheorem{dis:ass2}[dis:ass1]{Discussion}

\begin{abstract}
Wireless communication networks rely heavily on channel state information (CSI) to make informed decision for signal processing and network operations. However, the traditional CSI acquisition methods is facing many difficulties: pilot-aided channel training consumes a great deal of channel resources and reduces the opportunities for energy saving, while location-aided channel estimation suffers from inaccurate and insufficient location information. In this paper, we propose a novel \emph{channel learning} framework, which can tackle these difficulties by inferring unobservable CSI from the observable one. We formulate this framework theoretically and illustrate a special case in which the learnability of the unobservable CSI can be guaranteed. Possible applications of channel learning are then described, including cell selection in multi-tier networks, device discovery for device-to-device (D2D) communications, as well as end-to-end user association for load balancing. We also propose a neuron-network-based algorithm for the cell selection problem in multi-tier networks. The performance of this algorithm is evaluated using geometry-based stochastic channel model (GSCM). In settings with $5$ small cells, the average cell-selection accuracy is $73\%$ - only a $3.9\%$ loss compared with a location-aided algorithm which requires genuine location information.
\end{abstract}
\IEEEpeerreviewmaketitle

\section{Introduction}
Channel state information (CSI) plays a vital role in wireless communication systems. Both signal processing functions such as pre-coding and network operations such as user association requires CSI. The most prevalent method to obtain CSI is through pilot-aided channel training, in which a known sequence of signals is sent between the transmitter and the receiver so that the channel responses can be estimated. Nevertheless, the price is the channel resources and energy required to transmit and process pilots. This drawback renders pilot-aided channel training seriously challenged by the recent developments of wireless networks. Due to the recent surge in mobile traffic, cellular networks are now witnessing rapid densification\cite{WhatWill5GBe}. One part of this trend is the shrinking cell size and the increasing spatial reuse. The overlapping coverages of multiple access points (APs), which can be from the same tier, multiple heterogeneous tiers, or different radio access technologies (RATs), facilitate alternative user association strategies that are optimized for interference avoidance, load balancing, energy saving, and etc \cite{CHORUS}. Nevertheless, pilot-aided channel training requires the network to acquire the CSI between a massive amount of antenna pairs. Moreover, when base station (BS) sleeping is introduced into the network \cite{HCN}, the CSI about sleeping BSs will be missing unless they are frequently turned on to transmit pilots. This greatly reduces the energy saving opportunities. Another part of the network densification trend is contributed by massive {MIMO} \cite{Erick14}. The large number of antennas at BSs help the network to fully exploit the degrees of freedom (DoF) in spatial domain, making it possible to serve a large number of users simultaneous using the same piece of spectrum resources. However, the contamination of reused pilots in neighboring cells is shown to be the main limiting factor for massive {MIMO} systems \cite{pilot}, and downlink CSI is difficult and costly to obtain in frequency-division-duplexing ({FDD}) massive {MIMO} systems because of the large number of BS antennas and the expense of uplink feedback.

Realizing the drawbacks of traditional pilot-aided channel training, location-aided channel estimation has been proposed to infer the statistical CSI of users with the help of geographical channel database \cite{Locationaided}. Some recent work also utilizes the Guassian model to extrapolate channels for locations that are not registered in the database \cite{Locationaware}. However, accurate localization is generally expensive and difficult: network-side localization which utilizes the angle-of-arrival (AoA) and time-of-arrival (ToA) information is often inaccurate because of multi-path propagation and small system bandwidth; at the same time, device-side localization which rely on satellite assistance can drain device battery fast and is inapplicable to indoor scenarios. These deficiencies make location-aided methods only applicable for estimating coarse CSI in limited scenarios.

In this paper, we propose a novel \emph{channel learning} framework which can infer unobservable CSI from the observable one. Essentially, this framework is built upon the dependence between channel responses and location information. However, channel learning \emph{does not} explicitly exploit location information like existing location-assisted methods. Instead, this framework regards location information as a hidden factor and only deals with the relationship between observable and unobservable channel responses. Harnessing the power of machine learning models, we can build effective algorithms for approximating the mapping between the observable and unobservable channels. The channel learning framework is very flexible and can be applied into many scenarios, including cell selection in multi-tier networks, device discovery in device-to-device (D2D) communications, as well as end-to-end network operations.

The rest of the paper is organized as follows. In Section \ref{sec:framework}, we introduce the channel learning framework and discuss its practicability. In Section \ref{sec:app}, we describe the various applications of the channel learning framework. An algorithm that is based on the neuron network (NN) is then proposed for cell selection learning in multi-tier networks in Section \ref{sec:algorithm}. In Section \ref{sec:simulation}, We evaluate the accuracy of the proposed algorithm using geometry-based stochastic models and compare it with other algorithms. The paper is concluded in Section \ref{sec:conclusion}.

\section{Channel Learning Framework}\label{sec:framework}
Channel learning is the process of inferring unobservable channel information from the observable one. Assume a user is equipped with a single-antenna device to communicate with BS antennas in set $\mathbb{A}$. However, only the channel to the antennas which is in the observable subset $\mathbb{O} \subset \mathbb{A}$ can be obtained, whereas the channel to the rest of the antennas, which constitutes the unobservable subset $\mathbb{U} \subset \mathbb{A}$, is unknown to either the user or the network. We denote the channel\footnote{Although the direction of the link, i.e. uplink or downlink, does not change the channel learning framework, some configurations are more preferable due to practical consideration. For example, it is more convenient that the observable channels are all uplink channels so that the network can directly use them to infer unobservable channels, instead of having to collect feedback from user equipment.}
to the observable antennas as $\bm{h}_o$ and the channel to the unobservable antennas as $\bm{h}_u$. The user wants to know some metrics of the unobservable channel according to the metric function $\bm{m} = m(\bm{h}_u)$. Using these notations, we can formulate channel learning as the task to approximate a function $f(\cdot)$ from the observable channel to the unobservable metrics:
\begin{equation}\label{def:f}
\begin{aligned}
f \colon \mathbb{C}^{\vert \mathbb{O} \vert} &\to \mathbb{M} \\
\bm{h}_o &\mapsto \bm{m}
\end{aligned}
\end{equation}
where $\mathbb{C}$ is the set of complex numbers, $\vert \cdot \vert$ gives the cardinality of a set, and $\mathbb{M}$ is the co-domain of $m(\cdot)$. As long as function $f(\cdot)$ exist, we can utilize supervised learning algorithms to approximate it. Here we note that the training phase for supervised learning algorithms can be an off-line procedure. So the unobservable metric $\bm{m}$ in the training set can be easily calculated from the unobservable channel samples, which can be collected through traditional pilot-aided channel estimation during the training phase. Once training is finished, only the observable channel responses are needed to predict the unobservable metric.

The existence of $f(\cdot)$ is not trivial and cannot be guaranteed in the general sense. For example, in a propagation environment where there are no scatterers and just line-of-sight (LoS) propagation, users that lie on symmetric locations with respect to an uniform linear array will have the same array responses. But their channels to the antennas that do not lie on the symmetric axis could be different. If we regard the array response as the observable channel $\bm{h}_o$ and the channels to other antennas as $\bm{h}_u$, then one $\bm{h}_o$ may be mapped to two different values of $\bm{h}_u$ and $\bm{m}$. Thus $f(\cdot)$ does not exist. Therefore, we should carefully prove the existence of $f(\cdot)$ in different scenarios. In the next theorem, we provide a special scenario in which the existence of function $f(\cdot)$ can be guaranteed.
\begin{theo:exist}\label{theo:exist}
If the channel between the user and the BS antennas is a function of the user's geographical location $\bm{x}$, i.e. $\bm{h}_o = g_o(\bm{x})$ and $\bm{h}_u = g_u(\bm{x})$, and the function from user locations to observable channels $g_o(\cdot)$ is invertible, then there exist a function $f(\cdot)$ as defined in (\ref{def:f}).
\end{theo:exist}
\begin{proof}
If $g_o(\cdot)$ is invertible and its inverse is $g_o^{-1}(\cdot)$, the user location can be obtained from observable channel as $\bm{x} = g_o^{-1}(\bm{h}_o)$, and the unobservable channel can be derived as $\bm{h}_u = (g_u \ast g_o^{-1})(\bm{h}_o)$. Then the unobservable metric can be directly calculated as $\bm{m} = (m \ast g_u \ast g_o^{-1})(\bm{h}_o) = f(\bm{h}_o)$.
\end{proof}
Note even though $g_o(\cdot)$ can be invertible, given the difficulty to obtain accurate location information, we still cannot infer location $\bm{x}$ explicitly from the observable channel $\bm{h}_o$.

The applicability of the above theorem depends on two assumptions: 1) channel response is a function of user location; and 2) observable channel function $g_o(\cdot)$ is invertible. Next we discuss the practicability of these assumptions:
\begin{dis:ass1}[Location Determines Channel Responses]
	Channel responses in wireless fading channels can be decomposed into three factors: path loss, large-scale fading, and small-scale fading. Among them, path loss and large-scale fading is often considered to be static over time. Therefore, these two factors can be considered an unknown but deterministic function of user location. Small-scale fading is caused by the interference among multi-path components coming from ambient scattering. The relative location between the user, the scatterers, and the BS antennas should suffice to determine small-scale fading. As a consequence, small-scale fading will be a deterministic function of user location as long as the scatterers remain in the same position. Even if some of the scatterers do move around, the small-scale fading will only change negligibly over time if the scattering power from these scatters is small. In such cases, channel responses is still approximately a function of the user location.
\end{dis:ass1}
\begin{dis:ass2}[Invertible Observable Channel Function]
	The invertibility of the observable channel function $g_o(\cdot)$ can be guaranteed by the utilization of large number of antennas. In a rich-scattering environment with unlimited number of observable antennas, it is known that the channel responses from different users are asymptotically orthogonal \cite{marzetta-unlimited}. Thus, we know the observable channel responses from different users are asymptotically different:
	\begin{equation}\label{eq:neq}
	g_o(\bm{x}_1) \neq g_o(\bm{x}_2) \text{, if }  \bm{x}_1 \neq \bm{x}_2.
	\end{equation}
	Therefore, we can say that $g_o(\cdot)$ is an one-to-one mapping from user location to the observable channel responses, and $g_o(\cdot)$ is invertible. 
	
	Moreover, even if there are only a limited number of scatterers and observable antennas, the invertibility of $g_o(\cdot)$ can still be observed. Fig. \ref{fig:dist} illustrates such a numerical example using geometry-based stochastic channel model (GSCM). We randomly place $2$ users and $20$ scatterers within a semicircle with a radius of $700$m and calculate the channel responses from these two users to an uniform linear array placed at the center of this semicircle. The relationship between channel distances and users' geographical distances are shown in the figure, each blue cross representing the results of a random simulation run. As can be seen, the lower and upper bounds for channel distances stretch from the origin and diverge as we increase the distance between users. This implies that relationship as described in (\ref{eq:neq}) is guaranteed and the array response in this setting is an invertible function of user location.	
	\begin{figure}[!t]
		\centering
		\includegraphics[width=3in]{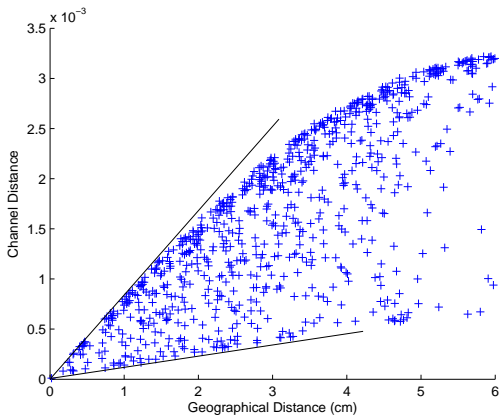}
		\caption{Channel distance v.s. geographical distance between two users. GSCM simulation using $20$ scatterers, Rician factor $K=10$ dB, uniform linear array with $100$ units.}
		\label{fig:dist}
	\end{figure}	
\end{dis:ass2}

\section{Applications of Channel Learning} \label{sec:app}
Channel learning is a flexible framework with a wide variety of potential applications. It is especially valuable for scenarios where it is costly or difficult to obtain channel information. In this section, we discuss some of these applications and describe how the framework is applied. 
\subsection{Cell Selection Learning in Multi-tier Networks}
Consider the problem of cell selection in multi-tier networks. The network needs to know the user's channel to a large number of small cells in order to select the best cell. This usually requires the network to sacrifice scarce channel resources for channel sounding. And users must scan for potential small cells, which is extremely energy- and time-consuming. Moreover, frequent channel sounding reduces the opportunity for BS sleeping in networks with dynamic sleeping operations, which greatly limits the potential for energy saving.

The channel learning framework can be applied here to solve these problems. A possible configuration is to place large antenna arrays at large cells and use them as the observable antennas to infer the channel to the overlapping small cells. The users intending to communication send sounding signals to the observable array at the large cell. Based on the array response $\bm{h}_o$, the network infers the best small cell for each user. Here the unobservable channel $\bm{h}_u$ is the channel between the user and small cells, and the unobservable metric function $m(\cdot)$ gives the index of the best small cell. In this case, the channel resources consumed for channel sounding is only proportional to the number of active users; the users just need to send sounding information and wait for the networks cell-selection suggestion; and the sleeping small cells that are not selected can continue to sleep.

\subsection{Device Discovery Learning in D2D Communications}
The discovery of proximate peer devices is a central problem in D2D communications. However, the traditional solutions either require devices to transmit and receive sounding signals, which greatly reduces battery life, or demand the network to provide location information, which is often inaccurate \cite{FodorD2D}.

The channel learning framework provides a new angle to tackle this problem. Specifically, we consider the case of network-assisted device discovery learning using massive BS antenna array. Assume we want to decide the proximity of two single-antenna devices at locations $\bm{x}_1 \in \mathbb{R}^3$ and $\bm{x}_2 \in \mathbb{R}^3$, respectively. Imagine a ``virtual'' device whose location is the composite of two actual device locations $\bm{x} = \{\bm{x}_1,\bm{x}_2 \} \in \mathbb{R}^6$. Then we can define the ``virtual'' observable channel $\bm{h}_o$ as the composite of the array responses from these two devices, the unobservable channel $\bm{h}_u$ to be the channel between these two devices, and the metric $m(\cdot)$ to be the probability that $\bm{h}_u$ is sufficiently good for direct communications.

Using the device discovery learning described above, the network does not need to estimate the actual device locations, which are inaccurate anyway. And the peering decisions can be determined in a centralized manner by the network, instead of in a distributed manner by the individual devices. This could greatly increase the spectrum and energy efficiency for both devices and the network.

\subsection{Extended Applications}
The channel learning framework can also be extended to predict channel metrics for users with multiple antennas, which is becoming a prevalent for smart devices such as mobile phones. In such a case, the metric function $m(\cdot)$ can be defined to reflect certian properties of the channel matrix, e.g. matrix rank or distribution of eigen-values. As discussed before, the feasibility of this extension is guaranteed as long as the assumptions in Theorem~\ref{theo:exist} is met.

Another potential extension for the channel learning framework is for making end-to-end network operations decisions which involve channel information. Take user association as an example. To make optimal user association decisions, the network needs to consider not only the channel quality between the users and the BS antennas so as to guarantee the quality of service (QoS), but also other factors such as the current load of each BS for load balancing . Extension in this aspect only requires a change in the metric function. Again consider user association for load balancing. To extend the channel learning framework, we allow the unobservable metric function to take in the BS loads $\bm{l}$ as an observable parameter, i.e. $\bm{m} = m(\bm{h}_u,\bm{l})$. In this way, the overall function $f(\cdot)$ is also extended to have two sets of arguments, i.e. $\bm{m} = f(\bm{h}_o,\bm{l})$. Following the derivation in in Theorem~\ref{theo:exist}, the existence of function $f(\cdot)$ can still be guaranteed if the two assumption are met.

\section{A NN-based Algorithm}\label{sec:algorithm}
In order to learn the mapping function $f(\cdot)$, we can employ supervised learning algorithms from the field of machine learning. Concretely, we first obtain a training set which contains some observations of $\bm{h}_o$ and the corresponding $\bm{m}$. Then we use these training data to choose a candidate function that best approximates function $f(\cdot)$. For illustration, we propose an NN-based algorithm for cell selection learning in multi-tier network. This algorithm combines the standard approaches in NN with special processing procedures that are tailored to the cell selection problem.

\subsection{Extracting Angular-Domain Information}
The first step of our algorithm is to extract angular-domain information from the raw observable channel response $\bm{h}_o$. We do this because it is hard for machine learning algorithms to automatically decode the complex relationship between spatial-domain channel response and user locations. In contrast, angular-domain information if more directly related with location information. So we have to manually convert the channel response into angular domain. Since the antennas of a large cell form a uniform linear array, we use the Fourier transform to obtain the angular-domain observable array response. After that, we extract the magnitude of the angular-domain response and omit the phase information.

\subsection{Forming Quantized Input Feature}
The second step is to translate the angular-domain magnitude into the feature space. It is well known that NN cannot handle input features that have large dynamic range. This is because the activation function of NN is the Sigmoid function, which saturates at extremely large or small input values. However, the dynamic range of the channel responses may span several orders of magnitude since users may lie in a wide geographical area. To tackle this problem, we take the logarithm of the angular-domain magnitude and then use the Lloyds algorithm to quantize it. After the logarithmic processing and quantization, the dynamic range of the input can be greatly reduced. The resulting input feature vector is formed by the quantization codebook index (normalized to $[-1,1]$) of the original angular-domain magnitude vector.

\subsection{Output Encoding}
The natural form for the output of cell selection learning is an integer index indicating the best small cell. However, due to the same reason as input encoding, the NN cannot well approximate a function with integer-valued output. For this reason, we employ the probabilistic output coding scheme. Suppose we need to choose from $K$ small cells, then the output is encoded as a length $K$ vector, with each entry stands for the probability that the corresponding small cell is the best cell. In this paper, we user a hard coding scheme for training sets. We set the entry that correspond to the best small cell to value $1$, and other entires to value $0$. However other soft-coding schemes that have multiple non-zero entries are also allowed in our algorithm. For example in the {SoftMax} encoding, each entry of the output vector can be set as the ratio between the channel magnitude of the corresponding antenna and the sum of the channel magnitudes of all the unobservable antennas.

\subsection{Cost Function and Training}
The cost function we use in the training phase is the sum of two parts. The first part is the average \emph{Kullback cross-entropy} \cite{cross-entropy} between the current NN outputs and the desired training outputs. Since the output can be interpreted as probability distribution, cross-entropy cost can be naturally applied. Also, cross-entropy cost function is known as a well formed cost function that provides better local minima than other cost functions such as squared error \cite{golik2013}. The second part of our cost function is the regularization term. Assume the NN has $N$ parameters $\{\theta_n, n = 1,2,\cdots,N-1,N\}$, the regularization term is defined as $$\lambda\sum\limits_{n=1}^{N}\theta_n^2,$$ where $\lambda \ge 0$ is the regularization parameter. Intuitively, proper regularization guarantees the ``sparsity'' of the parameters, which in turn avoids over-fitting. We employ the \emph{back-propagation} algorithm to derive the partial gradient of the cost function with respect to the network parameters $\{\theta_n, n = 1,2,\cdots,N-1,N\}$ and use \emph{conjugate gradient} algorithm to minimize the cost function. Note that other optimization algorithms can also be used to minimize the cost function.


\subsection{Cell Selection Prediction}
After the NN has been trained, we can use it to select the best small cell. We take the array response from a user and pre-process the array response to form the quantized input feature. Then we feed the input feature into the trained NN. The NN will output a selection probability for each of the output vector entry. We select the small cell with the largest selection probability. Note the entries of the output vector may not (and often don't) sum to probability one due to approximation error. This should not cause any problem in our application since only the relative magnitude of the entries matters. However, if the output probability distribution is to be used later, we will need to compensate the approximation error by means such as re-normalization.

\section{Simulation Results}\label{sec:simulation}
\begin{figure}[!t]
	\centering
	\includegraphics[width=3.7in]{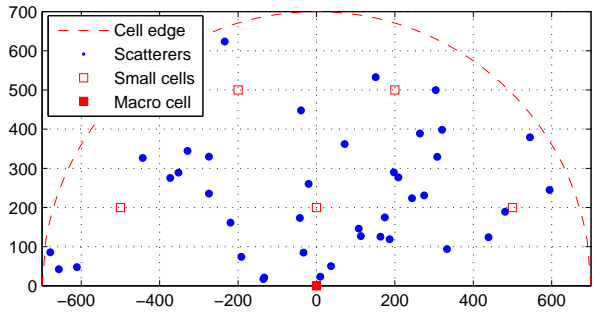}
	\caption{Network layout for simulation. The macro cell is the red filled square located at the origin, red hollow squares represent small cells, and blue dots represent scatterers.}
	\label{simu_setting}
\end{figure}
\begin{figure}[!t]
	\centering
	\includegraphics[width=3.7in]{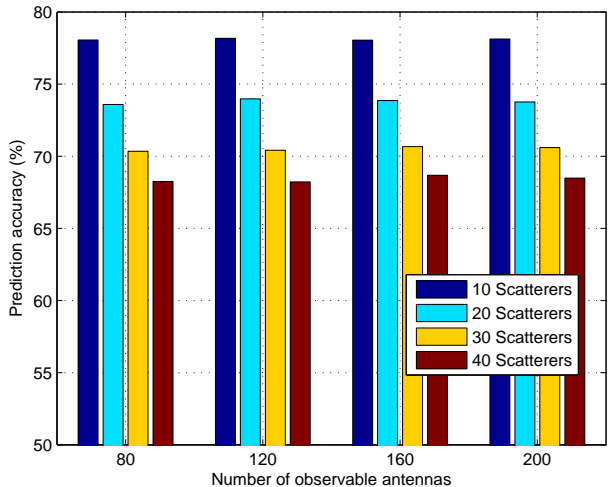}
	\caption{Prediction accuracy v.s. different number of antennas and scatterers.}
	\label{simu_result1}
\end{figure}
\begin{figure}
	\centering
	\includegraphics[width=3.7in]{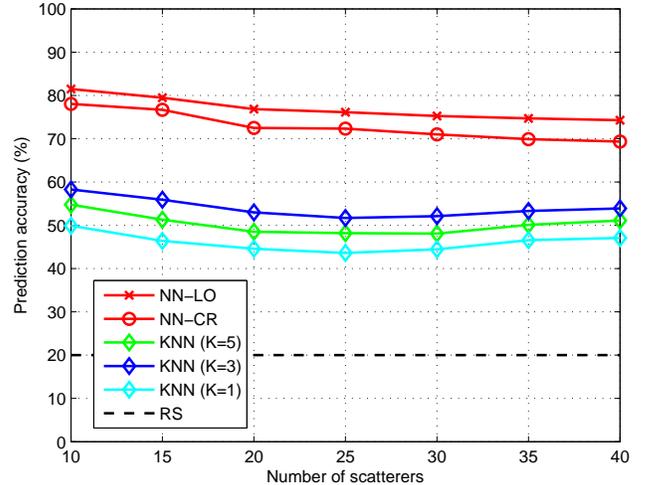}
	\caption{Prediction accuracy of different algorithms}
	\label{simu_result2}
\end{figure}
In this section, we evaluate the performance of the proposed learning algorithm in cell selection problems. We first introduce the simulation settings and then show the accuracy of the proposed algorithm.

\subsection{Simulation Settings}
We use the GSCM \cite{gscm} to generate fading channels in our simulation. The basic procedure is to firstly randomly place scatterers, calculate the complex channel gains of multi-path components, and then sum them up to get the final fading channel response. For simplicity, we only consider LoS propagation and single scattering.

The network layout is shown in Fig. \ref{simu_setting}. We consider a single macro cell and $5$ small cells. The coverage area of the macro cell is a circle with a radius of $700$m. To avoid mirroring ambiguity, we only consider the upper half of the circle. Scatterers are randomly dropped in the coverage area of the macro cell following uniform distribution. The macro cell is equipped with an uniform linear array with half-lambda antenna spacing. The Rician factor is 10 dB. And noise is ignored in the simulation.

In every simulation run, we randomly generate scatters and $2000$ user locations following uniform distribution, then randomly choose half of the samples for training and the rest half for testing. The prediction accuracy is defined as the percentage of the correct predictions on the test set. The final accuracy is the average accuracy over $50$ simulation runs.

\subsection{Simulation results}
Fig. \ref{simu_result1} shows the prediction accuracy (i.e. the percentage of predictions by which the strongest cell is correctly chosen) of the proposed algorithm with different number of observable antennas and scatterers. We can see that the accuracy slightly increases with the number of observable antennas. This is due to the improvements in angular-domain resolution with more antenna units in the array. Meanwhile, the prediction accuracy decreases with the number of scatterers. This is because the scattering environment becomes more complex when there are more scatterers, making the target function $f(\cdot)$ less smooth and harder to approximate. To solve this problem, deeper machine learning models such as deep NN can be used.

We also compare our algorithm with other prediction algorithms and the results are shown in Fig. \ref{simu_result2}. As can bee seen, the most naive \emph{random selection} (RS) method provides the baseline accuracy, which is $1/5 = 20\%$ for all scatterer configurations. The simple \emph{$K$-nearest-neighbor} (KNN) algorithm, which outputs the dominant choice among the $k$ nearest neighbors in the channel space, increases the accuracy to about $50\%$. There are around $4\%$ variation for different choices of $K$. In comparison, the accuracy of the proposed NN-based channel learning (NN-CR) algorithm is around $73\%$, which is far more better than RS and KNN. We also investigate a NN-based algorithm using genuine user location as inputs (NN-LO). Although genuine user location is difficult to obtain in real networks, the performance of this algorithm can be seen as an upper-bound for the performance of NN-based channel learning. We can see that there is an $3.9\%$ gap between the accuracies of NN-CR and NN-LO. This gap is essentially due to the information loss induced by inverting a not-perfectly-invertible observable channel function.

\section{Conclusion}\label{sec:conclusion}
In this paper, we propose the novel channel learning framework to infer unobservable CSI in wireless communication networks using observable one. We give conditions under which this framework is applicable and discuss the practicability of these conditions. We also discuss the potential applications of this framework, including cell selection in multi-tier networks, device discovery for D2D communications, and end-to-end user association for load balancing. For the case of cell selection, we also propose a NN-based supervised learning algorithm. The performance of this algorithm is evaluated using GSCM and compared with some other algorithms. Results show that the proposed algorithm can achieve around $73\%$ accuracy in settings with $5$ small cells, which is much better than RS and KNN and only $3.9\%$ less than the NN-based algorithm that utilizes genuine user geographical information.

\section*{Acknowledgment}
This work is sponsored in part by the {National Basic Research Program of China (973 Program: No. 2012CB316001)}, the {National Science Foundation of China (NSFC)} under grant No. 61201191, No. 61321061, No. 61401250, and No. 61461136004, and Hitachi Ltd.


\IEEEtriggeratref{2}
\bibliographystyle{IEEEtran}
\bibliography{IEEEabrv,CR}

\end{document}